\pdfoutput=1

\documentclass[aip,rsi,reprint,floatfix]{revtex4-2}

\usepackage{graphicx}
\usepackage{amsmath}
\usepackage{amssymb}
\usepackage{dcolumn}
\usepackage{bm}
\usepackage[utf8]{inputenc}
\usepackage[T1]{fontenc}
\usepackage{mathptmx}
\usepackage{tabularx}
\usepackage{colortbl}
\usepackage{booktabs}
\usepackage[table]{xcolor}

\begin{document}

\title[Design and operation of APEX-LD]{Design and operation of APEX-LD: a compact levitated dipole for the confinement of electron--positron pair plasmas}

\author{A. Card}
\email{alexander.card@ipp.mpg.de}
\affiliation{Max Planck Institute for Plasma Physics, 85748 Garching, Germany}
\affiliation{Technical University of Munich, 80333 Munich, Germany}

\author{M.R. Stoneking}
\affiliation{Lawrence University, Appleton, Wisconsin, 54911 USA}

\author{A. Deller}
\affiliation{Max Planck Institute for Plasma Physics, 85748 Garching, Germany}
\affiliation{University of California San Diego, La Jolla, California, 92093 USA}

\author{E.V. Stenson}
\affiliation{Max Planck Institute for Plasma Physics, 85748 Garching, Germany}

\date{\today}

\begin{abstract}

The objective of the APEX (A Positron--Electron eXperiment) project is to magnetically confine and study electron--positron pair plasmas. For this purpose, a levitated dipole trap (APEX-LD) has been constructed. The magnetically levitated, compact ($7.5$-cm radius), closed-loop, high-temperature superconducting (HTS) floating (F-)coil consists exclusively of a No-Insulation (NI) Rare-earth Barium Copper Oxide (ReBCO) winding pack, solder-potted in a gold-plated-copper case. A resealable in-vacuum cryostat facilitates cooling (via helium gas) and inductive charging of the F-coil. The $70$-minute preparation cycle reliably generates persistent currents of $\sim 60$~kA-turns and an axial magnetic flux density of $B_0 \approx 0.5$~T. We demonstrate levitation times in excess of three hours with a vertical stability of $\sigma_z<20$~µm. Despite being subjected to routine quenches (and occasional mechanical shocks), the F-coil has proven remarkably robust. We present the results of preliminary experiments with electrons, and outline the next steps for injecting positron bunches into the device.

\end{abstract}


\maketitle

\section{\label{sec:intro}Introduction}

Interest in electron--positron pair plasmas dates back nearly half a century. \cite{tsytovich_laboratory_1978} The mass symmetry of pair plasmas is predicted to give rise to unique wave dynamics and notable stability compared to magnetized electron--ion plasmas.\cite{helander_microstability_2014, horton_drift_1999}  Moreover, there is evidence for naturally occurring electron--positron plasmas in the vicinity of extreme astrophysical objects.\cite{siegert_positron_2016, siegert_positron_2023, chen_illuminating_2025} Recent efforts to produce electron--positron plasmas in the laboratory, using high-energy particle beams\cite{arrowsmith_laboratory_2024} or intense lasers,\cite{sarri_generation_2015} have yielded dense, quasineutral distributions of unmagnetized, relativistic pairs. Targeting a different parameter regime, the APEX (A Positron--Electron eXperiment) project\cite{pedersen_plans_2012} aims to create and confine low-temperature distributions of electrons and positrons and to investigate whether magnetized pair plasmas are as quiescent as predicted.\cite{stoneking_new_2020}

The first challenge to producing an electron--positron plasma is the limited availability of antimatter. Our target density ($n_e$) and temperature ($k_B T$) regimes are based on those that have already been achieved with pure-positron (non-neutral) plasmas in linear traps ($n_e \sim 10^{12}$~m$^{-3}$; $k_B T \sim 1$~eV).\cite{danielson_plasma_2015} The number of positrons required to produce a plasma scales with the device size, making smaller devices preferable, subject to limitations imposed by engineering, material, and experimental constraints. The reactor-based NEPOMUC (Neutron-induced Positron source MUniCh) facility\cite{hugenschmidt_nepomuc_2012} is capable of delivering $5 \times 10^7$ remoderated (i.e., low-energy, high-brightness) positrons per second.\cite{dickmann_upgrade_2020} It is conceivable, therefore, that $\sim 10^{10}$ positrons can be accumulated in a reasonable amount of time, sufficient for a confinement volume of $V \sim 10$~liters. Our target parameters correspond to a Debye length ($\lambda_D$, charge-screening length) of $\sim 1$~cm. As this is considerably shorter than the dimensions of the confinement volume ($V^{1/3} \sim 22$~cm), collective plasma effects are expected to influence the particle behavior.\cite{stenson_debye_2017}

The APEX levitated dipole (APEX-LD) has been designed to magnetically confine electron--positron pair plasmas.\cite{stoneking_new_2020} The magnetic field of the confinement volume is generated by current flowing in a superconducting coil that is magnetically levitated in vacuum. There are several engineering challenges associated with building a levitated dipole. The floating coil (F-coil) must be cooled to cryogenic temperatures and a persistent current must be generated in the closed loop of the superconductor. The F-coil must then be launched and stabilized, which can be accomplished using a concentric lifting coil (L-coil) to balance gravity. Earnshaw’s theorem states that there is no stable equilibrium in this configuration,\cite{earnshaw_nature_1842} thus the position of the F-coil must be continuously monitored and actively stabilized (e.g., by modulating the supply current of the L-coil).

Superconducting F-coils were first used for plasma studies in the levitated variants of the early ``internal ring'' devices, such as the Floating Multipole (FM-1) at PPPL. FM-1 explored toroidal field configurations such as the spherator, where the poloidal component of the field was supplied by a levitated coil.\cite{yoshikawa_experiments_1973} Inspired by the Voyager~2 discovery of high-$\beta$ plasma in the Jovian magnetosphere\cite{krimigis_hot_1979} (where $\beta$ is the ratio of plasma / magnetic pressures), Hasegawa proposed that a levitated dipole could be employed in a fusion reactor.\cite{hasegawa_dipole_1987, hasegawa_d-_1990} The Levitated Dipole Experiment (LDX) at MIT\cite{garnier_design_2006} and the Ring Trap devices (Mini-RT and RT-1) at the University of Tokyo\cite{mito_engineering_2003, ogawa_construction_2009} were built to investigate this concept. Experiments with these devices have demonstrated turbulence-driven inward transport of confined quasineutral electron--ion plasmas\cite{boxer_density_2008} and local plasma $\beta$ exceeding unity.\cite{nishiura_improved_2015} Self-organization of a pure-electron plasma into a magnetospheric vortex with a centrally peaked density profile was also observed in RT-1.\cite{yoshida_magnetospheric_2010}

\begin{figure*}
    \centering
    \includegraphics[width=0.75\linewidth]{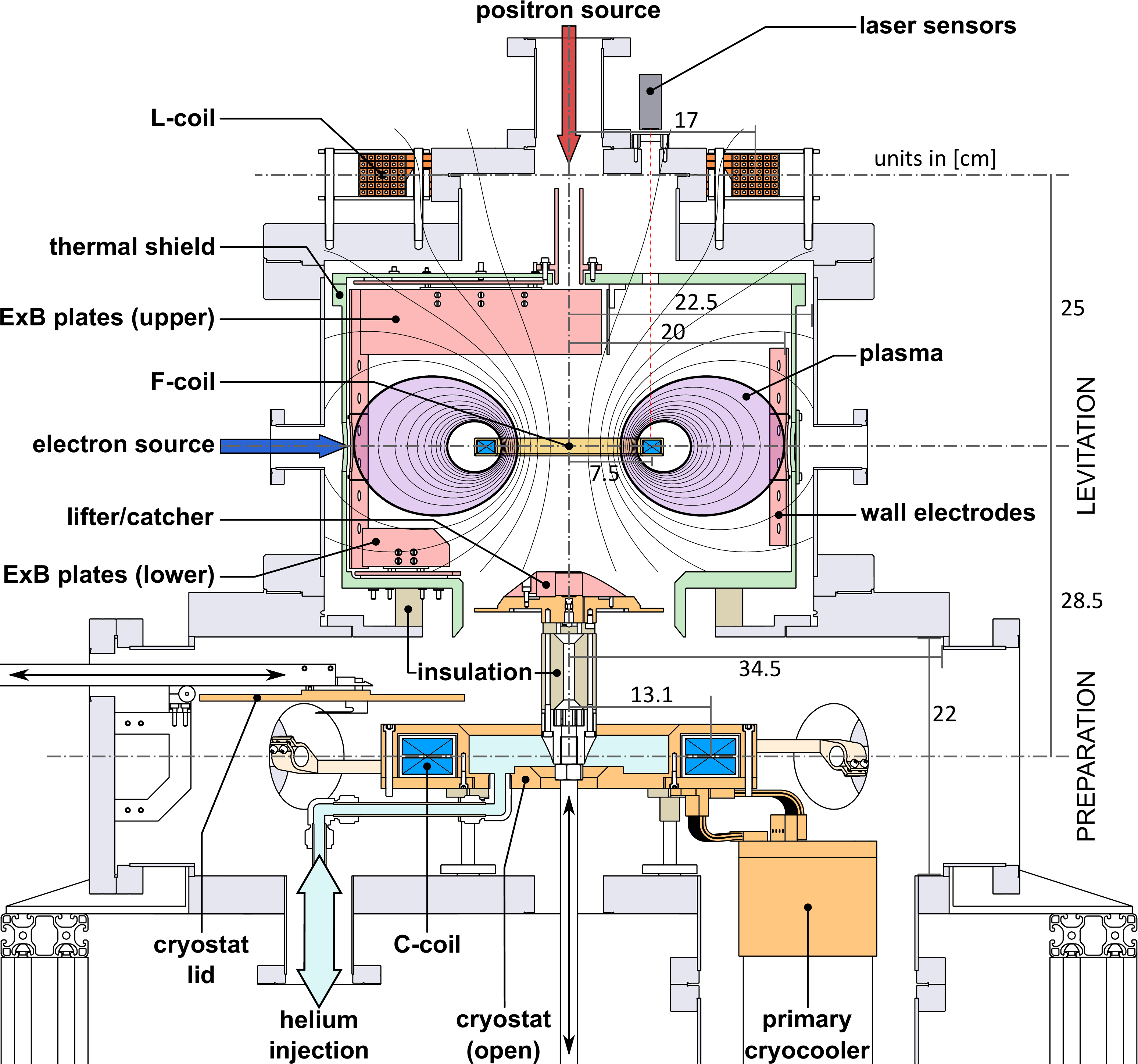}
    \caption{Schematic of APEX-LD during the levitation phase of operation. The upper levitation region hosts the F-coil (which produces the purple confinement region), L-coil, laser displacement sensors, thermal shield (green), and particle injection electrodes (pink). The lower preparation region hosts the cryostat (open, lid retracted), C-coil, primary cryocooler, horizontal and vertical translators (arrows indicate motion), and helium injection tubing. The lifter/catcher platform transports the F-coil between the two regions.}
    \label{fig:schematic}
\end{figure*}

In addition to the superconducting winding pack, the F-coil assemblies of previous devices have typically included an onboard cryostat with multi-layer insulation and demountable transfer lines for liquid or gaseous helium.\cite{file_operation_1971, zhukovsky_design_2000, smith_design_2001, ogawa_application_2006, mizumaki_development_2006} Energization of a persistent current was achieved either with a direct current (via detachable leads connected in parallel with a persistent current switch\cite{file_operation_1971, yanagi_excitation_2003, tosaka_development_2006}) or inductively using an additional charging coil (C-coil).\cite{zhukovsky_charging_2001} Flux pumps are currently being pioneered as an alternative approach by OpenStar Technologies in the private sector.\cite{chisholm_design_2025}  During levitation, the feedback-stabilized position of the F-coil was measured optically, either through occultation of a light source (using focused lamps\cite{foote_optical_1971} or lasers\cite{kesner_levitated_2003}) or with laser displacement sensors.\cite{morikawa_plasma_2004, morikawa_development_2007}

Our application constraints led us to make several design choices that distinguish APEX-LD from similar devices.\cite{card_compact_2025} The F-coil consists of a superconducting winding pack in a thin copper case to maximize the region of high magnetic field for plasma confinement. Contactless inductive charging within a resealable cooling enclosure allows us to eschew demountable connections. The correspondingly minimalist F-coil design takes advantage of recent developments with No-Insulation (NI) Rare-earth Barium Copper Oxide (ReBCO) high-temperature superconducting (HTS) materials, and fully soldered manufacturing techniques, to realize long levitation times and passive quench protection. We stabilize levitation of the F-coil to a precision of $\sigma_z \sim 20$~µm using feedback provided by a Field-Programmable Gate Array (FPGA) and real-time controller.\cite{card_fpga-stabilized_2024}

This paper is organized as follows. Section~\ref{sec:design} details the design of the system, with focus placed on the HTS coils (Sec.~\ref{sec:design:coils}), integrated cooling and charging station (Sec.~\ref{sec:design:cryostat}), and feedback-stabilized levitation system (Sec.~\ref{sec:design:levitation}). Section~\ref{sec:operation} describes the operation of APEX-LD, including the F-coil preparation (Sec.~\ref{sec:operation:preparation}) and levitation (Sec.~\ref{sec:operation:levitation}) phases, as well as a subsection on quenching and overall coil robustness (Sec.~\ref{sec:operation:quenching}). Section~\ref{sec:experiments} presents preliminary experiments with electrons that demonstrate the confinement capabilities of the device. In the final Section~\ref{sec:discussion}, we provide a discussion of future developments for APEX-LD.

\section{\label{sec:design}Design}

A schematic overview of APEX-LD is provided in Fig.~\ref{fig:schematic}. The ultra-high vacuum (UHV) chamber comprises two distinct regions. The lower preparation region contains a resealable cryostat (unsealed/open in Fig.~\ref{fig:schematic}), which cools the F-coil and encloses the C-coil used to induce a persistent current in its HTS windings. The upper levitation region contains the thermal shield, injection electrodes, and plasma diagnostics. The vertically translating lifter/catcher platform provides transport between the two regions. Fig.~\ref{fig:schematic} depicts APEX-LD during levitation. At the end of a levitation cycle, the F-coil is caught by the platform and returned to the preparation region.

\newcommand{\gap}{\hspace{1.5em}}
\setlength{\tabcolsep}{0pt}
\begin{table}
    \setlength{\tabcolsep}{-0.1pt}
    \rowcolors{2}{gray!15}{white}
    \centering
    \caption{APEX-LD parameters}
    \begin{tabularx}{\columnwidth}{
        >{\hsize=0.55\columnwidth\raggedright\arraybackslash}X
        >{\hsize=0.45\columnwidth\raggedleft\arraybackslash}X
    }
        \toprule
        \textbf{F-coil} & \\
        \gap Conductor material & NI GdBCO \\
        \gap Operating $T$ | $T_c$ & $20-75$~K | $92$~K \\
        \gap Average radius $r_F$ & $7.5$~cm \\
        \gap Winding pack $w_F \times h_F$ & $1.6$ $\times$ $1.3$~cm \\
        \gap Copper case $w_c \times h_c$ & $2.1$ $\times$ $1.7$~cm \\
        \gap Mass $m_F$ & $1.34$~kg \\
        \gap Self-inductance $L_F$ & $6.2$~mH \\
        \gap Resistance (@ $20$~K) $R_F$ & $72$~n$\Omega$ \\
        \gap Decay time constant $\tau=L_F/R_F$ & $24.1$~h \\
        \gap Persistent current $I_F \times N_F$ & $398$~A $\times$ $150$~t $= 59.7$~kA-t \\
        \gap Flux $\Phi_F$ & $16.5$~mWb \\
        \gap Flux density $B_{max}$ | axial $B_0$ & $1.2$~T | $0.5$~T \\
        \gap Stored energy $U_F$ & $493$~J \\
        \midrule
        \textbf{C-coil} & \\
        \gap Conductor material & GdBCO \\
        \gap Operating $T$ | $T_c$ & $25$~K | $92$~K \\
        \gap Average radius $r_C$ & $13.1$~cm \\
        \gap Winding pack | $w_C \times h_C$ & $4.6$ $\times$ $3.3$~cm \\
        \gap Vertical position* $z_C$ & $-28.5$~cm \\
        \gap Self-inductance $L_C$ & $56$~mH \\
        \gap Current $I_C \times N_C$ & $370$~A $\times$ $410$~t $= 151.7$~kA-t \\
        \gap Flux $\Phi_C$ & $50.3$~mWb \\
        \gap Stored energy $U_C$ & $3.4$~kJ \\
        \midrule
        \textbf{L-coil} & \\
        \gap Conductor material & Copper \\
        \gap Operating $T$ & $290$~K \\
        \gap Average radius $r_L$ & $17$~cm \\
        \gap Cross section | $w_L \times h_L$ & $4.4$ $\times$ $4.0$~cm \\
        \gap Vertical position* $z_L$ & $25$~cm \\
        \gap Current $I_L \times N_L$ & $80$~A $\times$ $36$~t $= 2.9$~kA-t \\
        \midrule
        \textbf{System} & \\
        \gap Cooldown ($300$~K$\rightarrow 20$~K) & $5.5$~h \\
        \gap Preparation & \\
        \gap \gap Initial cycle time & $70$~min \\
        \gap \gap Subsequent cycles & $45$~min \\
        \gap Levitation & \\
        \gap \gap Time & $3.25$~h \\
        \gap \gap Stability $\sigma_z$ & $18$~µm \\
        \gap Plasma confinement volume & $10$~liters \\
        \bottomrule
    \end{tabularx}
    \vspace{1ex}
    \parbox{0.795\textwidth}{\footnotesize *relative to levitated F-coil}
    \label{tab:parameters}
\end{table}

The plasma confinement volume is limited on the inboard side by the F-coil and on the outboard side by the wall electrodes at radius $r=20$~cm. The toroid formed by the uninterrupted field lines of the levitating coil has a volume of $10$~liters. The electrodes and thermal shield have not yet been installed and are part of a planned upgrade that will be discussed in Sec.~\ref{sec:discussion}. Table~\ref{tab:parameters} presents key physical and operational parameters for APEX-LD.

\begin{figure}
  \centerline{\includegraphics[width=0.8\linewidth]{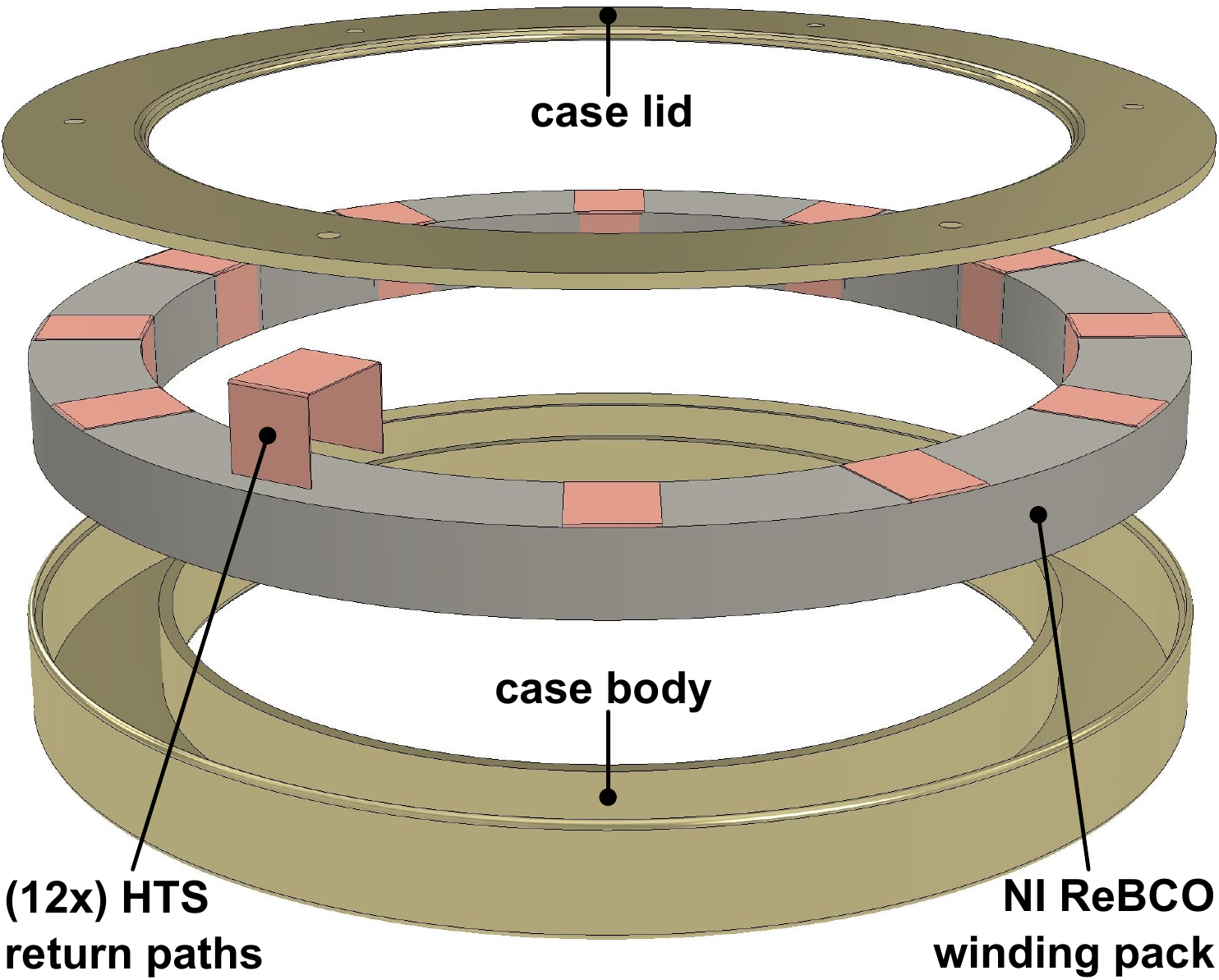}}
  \centerline{\includegraphics[width=0.8\linewidth]{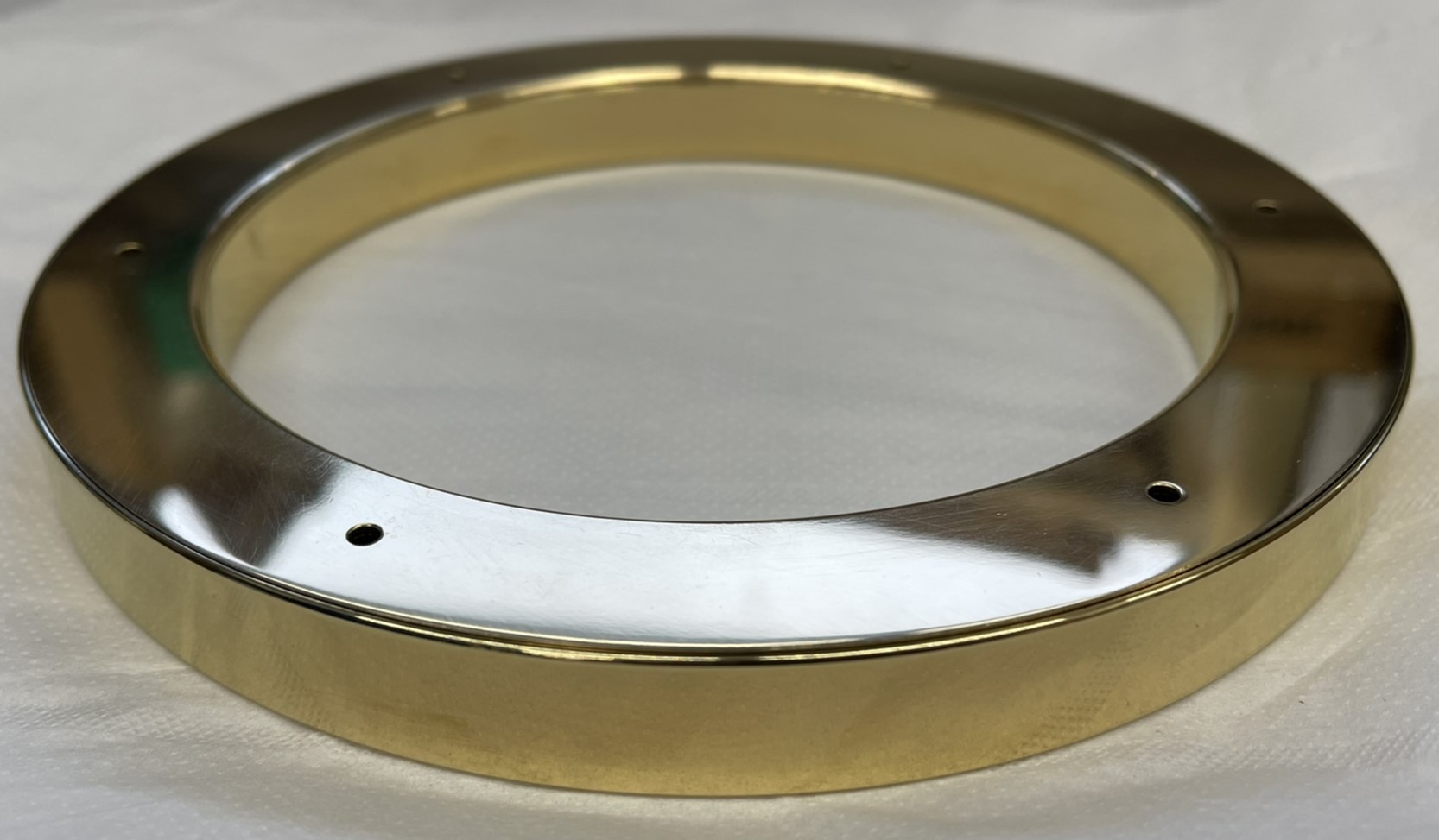}}
  \caption{(top) Exploded schematic of the F-coil. The NI ReBCO HTS winding pack is a single-pancake coil of $150$~turns closed by twelve radial bands connected in parallel. The fully soldered winding pack is soldered into a gold-plated copper case. (bottom) A photograph of the manufactured F-coil.}
  \label{fig:Fcoil}
\end{figure}

\subsection{High-temperature superconducting coils\label{sec:design:coils}}

\begin{figure*}
    \centering
    \includegraphics[width=1\linewidth]{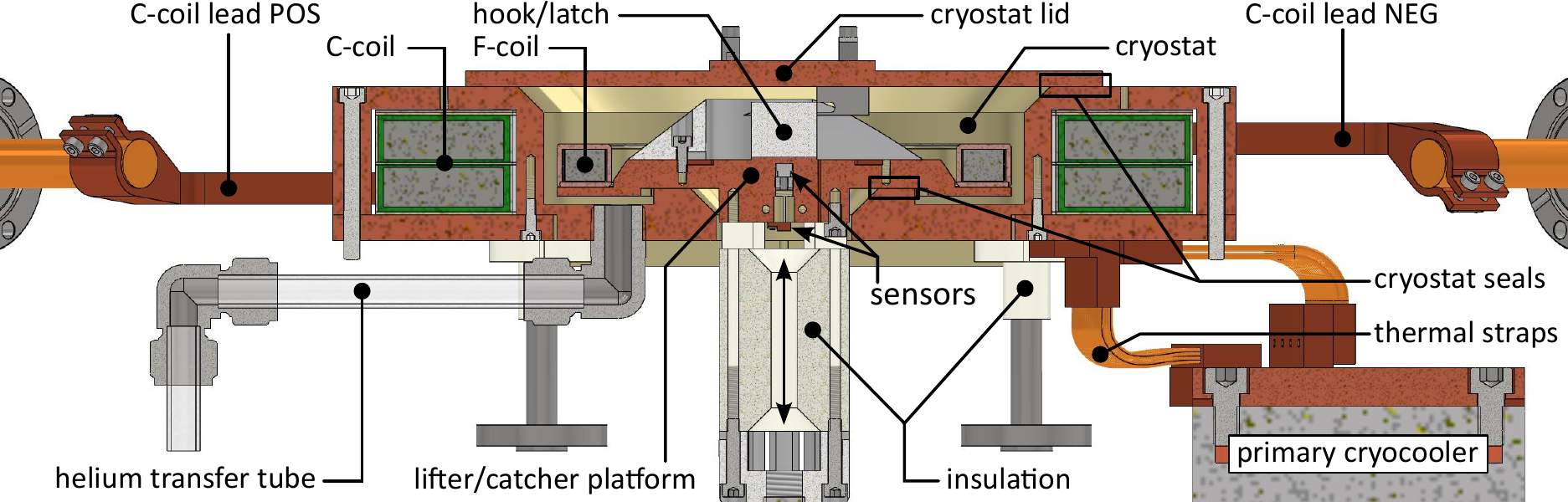}
    \caption{Cross section of the cryostat CAD model, presented in the closed (sealed) configuration for F-coil preparation. Note the coplanar, concentric alignment of the F-coil relative to the C-coil for good inductive coupling.}
    \label{fig:cryostat}
\end{figure*}

APEX-LD features two HTS coils (F-coil and C-coil) that were engineered in close collaboration with THEVA GmbH. While the two coils differ in design (e.g., single versus double pancake), they are both manufactured using ReBCO HTS tape produced by THEVA. The specific HTS compound is Gadolinium Barium Copper Oxide (GdBCO), which has a superconducting transition at $T_c \approx 92$~K.\cite{shi_superconducting_2008} The tape is composed of a thin superconductor layer ($\sim 5$~µm) deposited onto a magnesium oxide-coated Hastelloy substrate, which is then silver plated.\footnote{Since the Hastelloy substrate has relatively poor electrical and thermal conductivities, there is a preferred side to the tape for electrical and thermal connections.} Copper electroplating brings the thickness of the non-insulated tape utilized by the F-coil to $120$~µm. The polyimide-insulated tape used for the C-coil also includes an extra layer of copper laminate and has an overall thickness of $220$~µm. Both coils were constructed using 12-mm-wide tape (minimum normal bending radius of $3$~cm).

Figure~\ref{fig:Fcoil} presents an exploded view of the F-coil design and a photograph of the manufactured coil. It is a closed coil with no external leads nor persistent current switch. The coil must be inductively energized and resilient to routine quenches. At the time of manufacture, only $12$-mm-wide ReBCO tape was accessible, so a single-pancake design was chosen.  The winding pack consists of $150$~turns with an average radius of $7.5$~cm. Twelve U-shaped strips of ReBCO connect the inner and outer turns, providing a low-resistance parallelized return path for persistent current. The F-coil has a mass of $1.34$~kg and a square cross-sectional area of $3.6$~cm$^2$.

The F-coil NI winding pack is stabilized with tin-silver-copper solder (Sn96.5–Ag3.0–Cu0.5) and potted with indium-silver solder (In97-Ag3) inside a thin, gold-plated, oxygen-free copper (Cu-OF) case.  The fully soldered coil is mechanically, thermally, and electrically contiguous.  The low-emissivity case shields the winding pack from thermal radiation.  Its thin profile minimizes shadowing of the poloidal magnetic field lines closest to the coil.

The C-coil is used to inductively charge the F-coil.  It comprises $410$~turns of $12$-mm-wide, turn-to-turn insulated HTS tape, in a double-pancake configuration with the open ends soldered to copper electrodes. The average radius of the C-coil is $13.1$~cm, with winding pack dimensions of $46$~mm (horizontal) by $33$~mm (vertical). It is stabilized with a low out-gassing epoxy resin and sealed in the gold-plated Cu-OF enclosure of the cryostat (Sec.~\ref{sec:design:cryostat}).

\subsection{\label{sec:design:cryostat}Integrated cooling and charging station}

The F-coil is cooled and charged in a resealable cryostat located in the lower preparation chamber. Figure~\ref{fig:schematic} presents the cryostat in the open (unsealed) configuration employed during levitation. The closed (sealed) configuration is presented in Fig.~\ref{fig:cryostat}. At the center of the cryostat is a cavity into which the F-coil is retracted. The coplanar, concentric alignment between the two coils maximizes inductive coupling. A horizontal translator transfers the cryostat lid to the lifting platform using two sets of hooks and latches. Fully retracting the lifter engages two annular metal-to-metal seals, isolating the cavity volume from the rest of the vacuum chamber. The gold-plated-copper seals also provide good thermal contact, which ensures the lid and lifter rapidly thermalize with the cryostat body.

The primary cryocooler (Cryomech AL630) is connected to the cryostat with four flexible Cu-OF thermal straps, which provide vibration isolation and a thermal conductance of $6.7$~W/K at $20$~K. Thermal conductivity between the F-coil and cryostat is established through i) direct contact with the lifter, ii) blackbody radiation, and iii) an exchange gas (helium), which is transferred to and vented from the cryostat via a plastic (PFA) tube. With the cryostat lid sealed, helium can be introduced using a stepper-motor-controlled leak valve. In combination with a turbomolecular pump (Leybold TURBOVAC MAG W 1300 iP), the metal seals allow the cryostat to be pressurized to $\sim 1$~mbar with helium while maintaining a pressure of $< 1 \times 10^{-4}$~mbar in the vacuum system. The cryostat is supported on four stainless steel legs that are thermally insulated from the vacuum chamber by TECASINT 2011 standoffs (see Fig.~\ref{fig:cryostat}). From room temperature, the station cools to its base temperature of $20$~K in $5.5$~h, with the F-coil trailing by $1$~h (without an exchange gas).

Magnetic field strength is measured using a Hall sensor embedded in the lifter assembly (centered and coplanar with the two HTS coils).  The lifter platform is thermally and electrically isolated from the vertical translator using a TECASINT 2011 insulation standoff. This allows the lifter to serve as a plasma diagnostic (capacitive probe or charge collector), or it can be biased to assist with charged particle injection.

The C-coil enclosure is cooled by the primary cryocooler. Two pieces of single-crystal sapphire (Al$_2$O$_3$) ensure good thermal contact between the enclosure and C-coil electrodes without shorting the latter to ground. The cryogenic surfaces act as a cryopump that adsorbs impurities outgassing from the stabilizing epoxy. Current is supplied to the C-coil through two actively cooled leads (POS and NEG in Fig.~\ref{fig:cryostat}).  Each lead consists of a $270$-mm Cu-OF busbar that thermally connects a room-temperature vacuum feedthrough to the secondary cryocooler (Leybold COOLPOWER 250 MD), followed by two parallel strips of $12$-mm-wide NI GbBCO HTS tape that connect to the C-coil electrodes. The $160$-mm-long HTS bridges minimize the ohmic and conductive heat load to the C-coil (and provides vibration isolation). Similarly, the cross-sectional area of the Cu-OF busbars ($12\times3$~mm) is a compromise of thermal and electrical conductivity that minimizes the heat load to the secondary cooler during C-coil energization. On one end, the HTS strips are soldered to each busbar with a $\sim 70$-mm overlap and the connections are clamped to the secondary cryocooler through sapphire insulators, reaching thermal equilibrium at $31$~K ($32$~K when the C-coil is energized). On the opposite end, each HTS strip is soldered with a $\sim 20$-mm overlap to a Cu-OF adapter that clamps to the C-coil electrode. The C-coil power supply (Cryogenic SMS450C 4Q T R/M) provides up to $400$~A of current with built-in quench protection.  The maximum current that we can continuously supply to the C-coil is limited to $\sim 370$~A ($152$~kA-t) by the HTS bridges.  When the C-coil is energized, the lead assemblies contribute $4$~W of heat to the C-coil and $50$~W to the secondary cryocooler.

\subsection{\label{sec:design:levitation}Feedback-stabilized levitation control}

\begin{figure}
    \centering
    \includegraphics[width=1\linewidth]{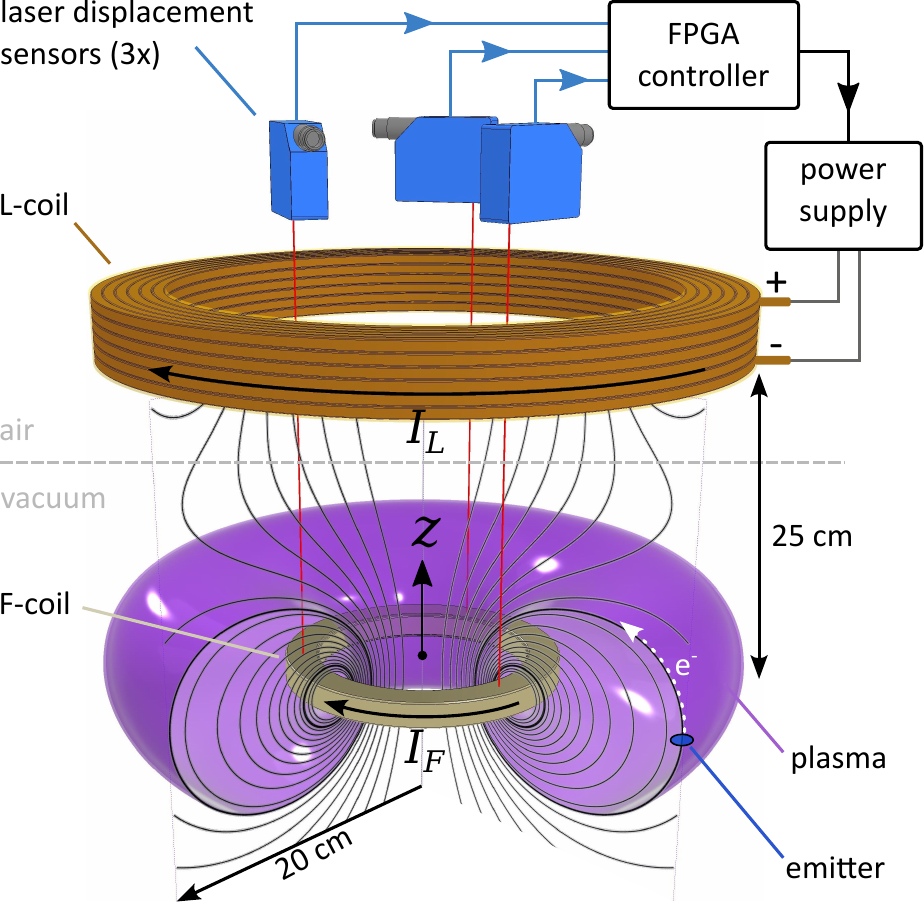}
    \caption{The feedback-stabilized levitation system. The larger diameter L-coil, located above, naturally stabilizes the slide and tilt motions of the F-coil. The feedback loop stabilizes the vertical motion ($\hat{z}$). An electron emitter can be inserted onto outboard field lines to inject electrons into the confinement volume.}
    \label{fig:levsys}
\end{figure}

The energized F-coil is magnetically levitated using the L-coil located above the vacuum chamber.  The axisymmetric coil geometry is configured to be inherently stable for five of the six degrees of freedom: slide ($xy$), tilt (pitch, roll), and azimuthal (yaw) rotation.\cite{saitoh_levitated_2020}  The vertical ($z$) position, however, must be actively stabilized.   Our stabilization system is based on digital PID feedback implemented by an FPGA-based controller.\cite{card_fpga-stabilized_2024} The software was developed using a test levitation system and a neodymium permanent magnet.\cite{saitoh_levitated_2020}

The physical components of the levitation system are presented in Fig.~\ref{fig:levsys}. The F-coil is located inside the vacuum chamber, whereas the laser displacement sensors, FPGA-controller, L-coil, and its power supply (Regatron TopCon Quadro\footnote{Recently replaced with a Heinzinger Power Converter 50-200.}) are all located outside of the vacuum system. The L-coil is wound from square copper profile with a central channel for water cooling. The $6 \times 6$ turn winding pack has an average radius of $17$~cm. The L-coil is positioned $25$~cm above the levitation plane and can be continuously driven with up to $200$~A of DC current.

The stabilization flow diagram is shown in Fig.~\ref{fig:levsys}. Three laser displacement sensors measure the vertical position of the F-coil.  The FPGA controller digitizes the sensor signals, rejects any out-of-range values (e.g., due to sensor blinding), and then feeds the average into a PID algorithm.  The digital PID output is then post-processed (e.g., range coercion) and converted to an analog output signal that manipulates the L-coil power supply current to correct displacements from the setpoint position. The magnetic field of the L-coil represents only a small (and axisymmetric) perturbation to the confinement field ($2.9$~kA-t in the L-coil versus $60$~kA-t in the F-coil).

\section{\label{sec:operation}Operation}

\subsection{\label{sec:operation:preparation}Preparation}

\begin{figure}
    \centering
    \includegraphics[width=1\linewidth]{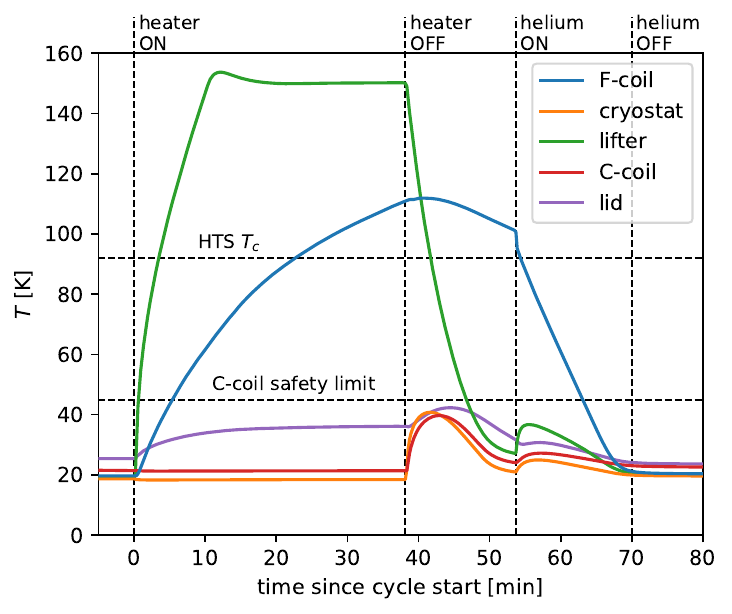}
    \caption{Temperature of the cryostat and HTS coils during F-coil preparation. The F-coil is heated by the lifter above $T_c$ and then rapidly re-cooled using helium gas.  The C-coil is kept below $45$~K so it can safely energized during the charging procedure.}
    \label{fig:chargingT}
\end{figure}

The inductive charging procedure thermally cycles the F-coil above and below its critical temperature, $T_c \approx  92$~K while keeping the C-coil below $45$~K (the limit for safely energizing with $400$~A). Measurements of the cryostat and F-coil temperatures during preparation are presented in Fig.~\ref{fig:chargingT}. At $t=0$, the lifter is raised by $1.5$~mm to thermally disconnect it from the cryostat. The F-coil is then warmed above its superconducting transition using a resistive heater embedded in the lifter and a PID feedback controller (Lake Shore Cryotronics Model 336) with a $150$-K setpoint.  After $38$~minutes, the F-coil has warmed to $110$~K, at which point the heater is disengaged and the cryostat is resealed. At $t=54$~minutes, the C-coil is turned on and the F-coil is re-cooled by helium gas, which is introduced into the cryostat via a stepper-motor-controlled leak valve connected to the transfer tube.  This raises the pressure in the cryostat to $\sim 1$~mbar and establishes a strong thermal connection between the F-coil and cryostat. At $t=70$~minutes, the F-coil has reached its base temperature and the helium is vented.

\begin{figure}
    \centering
    \includegraphics[width=1\linewidth]{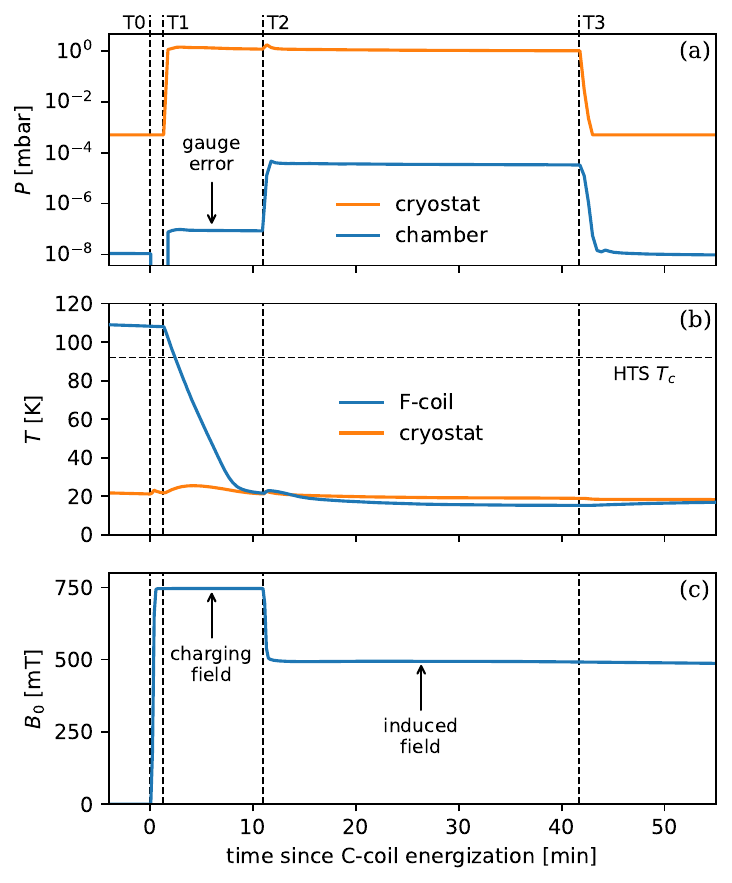}
    \caption{(a) Pressure, (b) temperature, and (c) axial magnetic flux density during the first F-coil induction test. Prior to the start, the F-coil was warmed to a normally conducting state. At $T0$ the C-coil is ramped on ($750$~mT). Helium is injected at $T1$ to cool the F-coil below the superconducting transition. At $T2$ the C-coil is ramped off, inducing a persistent current in the F-coil. Helium is vented at $T3$, and the preparation phase is complete.}
    \label{fig:induction}
\end{figure}

Figure~\ref{fig:induction} presents the pressure $P$, temperature $T$, and axial magnetic flux density $B_0$ during the inductive charging procedure. At $T0$, the F-coil is not superconducting.  The C-coil is ramped to $152$~kA-t over $10$~s, which generates a magnetic field of $750$~mT at its center.  The drop in the measured chamber pressure is caused by the stray magnetic field at the ion gauge. At $T1$, helium begins to cool the F-coil (i.e, $t = 54$~minutes in Fig.~\ref{fig:chargingT}) and at $T2$ it has reached its base temperature.  This particular measurement was taken before the HTS winding pack was soldered into the copper case, resulting in a slightly faster cooldown than shown in Fig.~\ref{fig:chargingT}.  With the F-coil now superconducting, the C-coil can be ramped off, which induces a persistent current of $I_F \approx 400$~A in the F-coil (i.e., $N_F I_F = 60$~kA-t; $B_0 = 500$~mT).  This indicates that the F-coil captures a sizable portion of the C-coil flux. Normally, the helium is vented immediately after charging the F-coil. However, for the measurement shown in Fig.~\ref{fig:induction}, the pressure was maintained for $30$~minutes for testing purposes. The large pressure differential between the inside and outside of the cryostat ($P_\mathrm{cryostat} \approx 10^4 \times P_\mathrm{chamber}$) allows the F-coil to be rapidly cooled without warming the cryostat / cooling the vacuum vessel.

\begin{figure}
  \centerline{\includegraphics[width=1\linewidth]{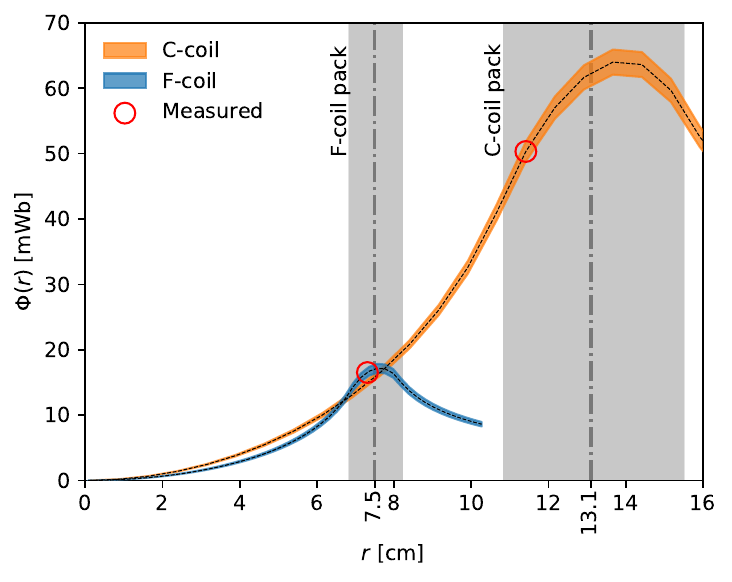}}
  \caption{Radially integrated flux as a function of $r$ for the C-coil (with $152$~kA-t) and the F-coil (with $60$~kA-t) at $z=0$. For each curve, the shaded extent along the ordinate represents the uncertainty. Intersection of measured flux values with the calculated curves are circled in red for each coil.}
  \label{fig:FluxMatch}
\end{figure}

The total magnetic flux $\Phi$, through a surface $S$, is given by the surface integral of the normal component of the magnetic flux density $\mathbf{B}$.  Assuming rotational symmetry about the $z$-axis and a circular surface with radius $r$ at a height $z=0$, this reduces to
\begin{equation}\label{eq:Bflux}
    \Phi (r) = \iint_{S(r)} \mathbf{B}(r,z) \cdot d\mathbf{S} = \int_{0}^{r} B_z(\rho,0) 2\pi \rho d\rho \; .
\end{equation}
\noindent 
Applying the Biot-Savart law to a infinitesimally thin, current-carrying loop of radius $a$ provides an analytical expression for the $z$ component of the off-axis magnetic field $B_z(\rho,z)$, in the plane of the loop ($z=0$),
\begin{equation}\label{eq:B_B-S}
B_z(\rho,0) = \frac{\mu_0 I}{2\pi (a+\rho)}\left[ K(k^2) + \frac{(a^2-\rho^2)}{(a-\rho)^2} E(k^2) \right] \; ,
\end{equation}
where $K$ and $E$ are complete elliptical integrals of the first and second kind,\cite{simpson_simple_2001} and
\begin{equation}\label{eq:B_B-S_k}
k^2 = \frac{4a\rho}{(a+\rho)^2} \; .
\end{equation}
The total flux enclosed by a coil with a finite winding pack can be determined from the superposition of the flux generated by an offset stack of concentric current loops.

Figure~\ref{fig:FluxMatch} shows the calculated enclosed flux as a function of $r$ (in the symmetry plane of the coils) for the fully energized C-coil, $\Phi_C(r)$.  Also shown is the equivalent calculation for the inductively charged F-coil, $\Phi_F(r)$. The radial extent of each winding pack is shaded gray. The flux of each coil was independently estimated by integrating the measured back electromotive force (EMF) as the C-coil was ramped on (F-coil not superconducting) and off (F-coil superconducting).  The intersection of the measured values [$\Phi_F(r_F)=16.5$~mWb; $\Phi_C(r_C)=50.3$~mWb] and the calculated curves are circled in red. The two curves closely match at the location of the F-coil, $\Phi_F(r_F) \approx \Phi_C(r_F)$, which suggests that all of the geometrically available flux is captured.

\begin{figure}
    \centering
    \includegraphics[width=1\linewidth]{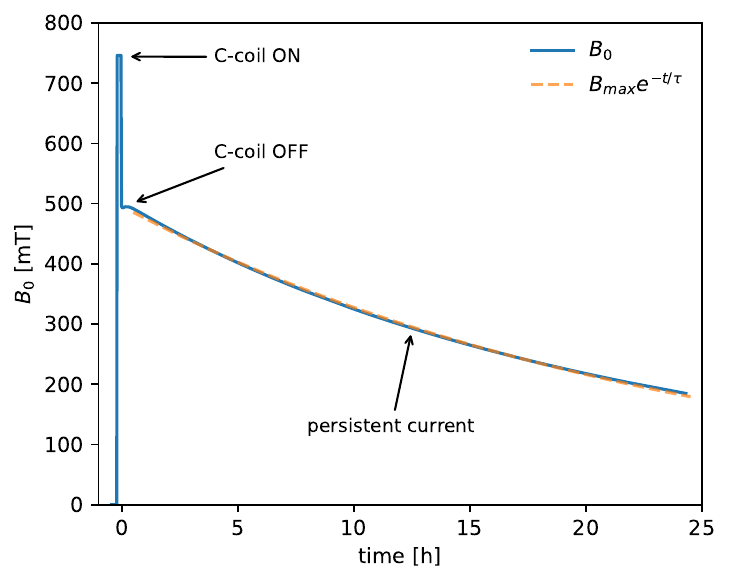}
    \caption{Slow decay of the axial magnetic field $B_0$ (measured at the center of the F-coil) when the charged F-coil is maintained at $T=20$~K (i.e., the continuation of Fig.~\ref{fig:induction}c). The measured decay time is $\tau=24.1$~h.}
    \label{fig:Bdecay}
\end{figure}

To measure the internal resistance of the F-coil, we observe the $L/R$ decay of its persistent current with the coil maintained at base temperature, as presented in Fig.~\ref{fig:Bdecay}. Fitting the exponential decay to $B_{max}e^{-t/\tau}$ indicates a decay time of $\tau=24.1$~h.  The self-inductance of the F-coil is calculated from its geometry to be $L_F = N_F \Phi_F / I_F = 6.2 \pm 0.3$~mH. Therefore, the residual resistance of the solder joints at $20$~K is  estimated to be $R_F=72 \pm 4$~n$\Omega$.

\subsection{\label{sec:operation:levitation}Levitation}

Upon completion of the preparation phase, the cryostat is opened using the two linear translators. The lifter/catcher platform, attached to the vertical translator, includes a latch that couples with cantilevered hooks on the underside of the cryostat lid (see Fig.~\ref{fig:cryostat}). A horizontal translator arm, featuring a trident hook, enters from the side and engages with three latches on top of the lid. The horizontal arm decouples the lid from the vertical translator and then retracts it out of the way (see Fig.~\ref{fig:schematic}). With the cryostat opened, the lifting platform is clear to lift the F-coil into the launching position.  At this point, the FPGA-based levitation controller and L-coil supply are switched on. As the L-coil lifts the F-coil, eddy currents are induced in the copper lifting platform.  This causes the launching coil to tilt,  resulting in temporary failures of the displacement sensors.  The associated error signals are automatically handled by the controller software, allowing it to continue the launch and then stabilize the position of the F-coil.\cite{card_fpga-stabilized_2024}

\begin{figure}
    \centering
    \includegraphics[width=1\linewidth]{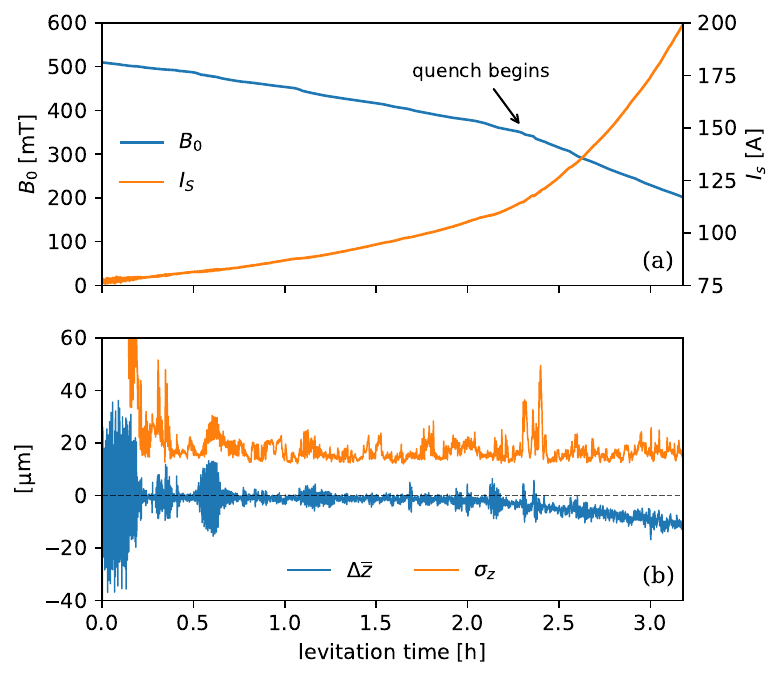}
    \caption{An example of a full levitation cycle. (a) Axial magnetic flux density $B_0$ and L-coil supply current $I_s$ required to maintain the F-coil setpoint position. (b) The average vertical displacement $\Delta \overline{z}$ and stability $\sigma_z$ of the F-coil over a $10$-s rolling window.}
    \label{fig:levcycle}
\end{figure}

Figure~\ref{fig:levcycle} presents an example levitation cycle. Figure~\ref{fig:levcycle}a shows the axial flux density $B_0$ at the center of the F-coil (calculated from the lifter Hall sensor and laser position measurements) and manipulated L-coil supply current $I_s$ (approximately equal to $I_L$) required to maintain the setpoint position $z_F$. Figure~\ref{fig:levcycle}b shows the average vertical displacement $\Delta \overline{z}$ and standard deviation of the averaged laser sensor signal $\sigma_z$ (i.e., the vertical stability), both taken over a $10$-s rolling window (with an acquisition rate of $100$~Hz). The total levitation time is $3.25$~h, with a vertical stability of $\sigma_z=18$~µm, calculated over the full cycle.\cite{card_fpga-stabilized_2024}

\subsection{\label{sec:operation:quenching}Quenching}

The NI ReBCO construction of the APEX-LD F-coil has exceeded expectations in terms of robustness. In Fig.~\ref{fig:levcycle}a, we see the point at which the quench begins. During levitation, thermal radiation gradually warms the F-coil, which increases the resistance of the current path and increases the rate of decay.  As the coil continues to warm, the critical current threshold also steadily decreases. This continues until the circulating current matches the critical current of the HTS tape, which provokes a notable increase in the decay rate. Since the current is not driven, any excess current is diverted through the resistive layers of copper and solder.  This threshold quenching process dissipates the energy stored in the magnetic field very slowly, which allows the F-coil to continue levitating for an additional $50$~minutes.

The magnetic field energy of the fully-charged F-coil is $U_F \approx 500$~J. At the onset of the quench, the current in the F-coil has resistively decayed, resulting in a reduction of $B_0$ to $0.7 B_{max}$ ($350$~mT). During the following $50$-min quenching period $B_0$ further reduces to $0.4 B_{max}$ ($200$~mT). Assuming the self-inductance ($L$) is constant, and
\begin{equation}\label{eq:Uenergy}
    U = \frac{1}{2} L I^2 \Rightarrow U \propto B_0^2 \; ,
\end{equation}
\noindent the reduction of $B_0$ by $\Delta B_{quench} = 0.3 B_{max}$ results in a conversion of $\Delta U_{quench} = 0.09 U_{max} \approx 36$~J of magnetic energy into heat energy ($Q$). Using the mass ($m$) of the F-coil, and specific heat capacity ($C$) of Cu-OF copper ($RRR=100$; $60$~K), we calculate a F-coil temperature rise using
\begin{equation}\label{eq:Qenergy}
    \Delta T = \frac{Q}{m C} \; .
\end{equation}
\noindent Inserting values into Eq.~\ref{eq:Qenergy} predicts an increase of $\Delta T \approx 0.7$~K over the $50$-min quench. This excess heat is easily absorbed by the solid mass of the F-coil, and corroborates the observation that when a thermal sensor was installed during commissioning, we did not see a measurable change in the F-coil temperature during the quench.

APEX-LD has performed over $30$ levitation cycles, all of which ended in an F-coil quench. Furthermore, the coil has endured a mechanical impact against the ceiling of the vacuum chamber at an estimated velocity of $5$~m/s.  Nevertheless, we have been unable to measure any degradation in performance of the fully soldered HTS coil.

\section{\label{sec:experiments}Electron experiments}

\subsection{\label{sec:experiments:visualization}Field line visualization}

\begin{figure}
    \centering
    \includegraphics[width=1\linewidth]{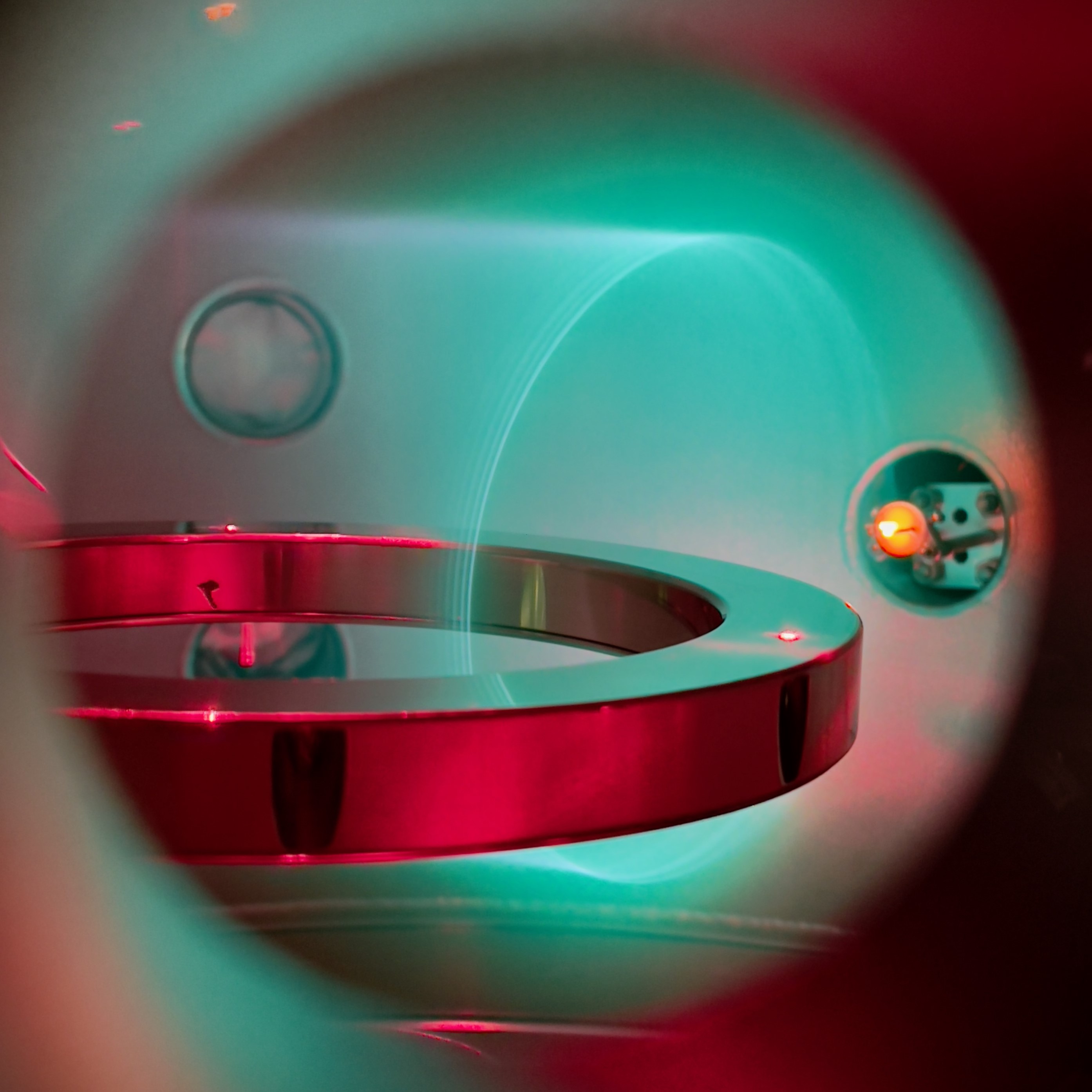}
    \caption{Photograph of the levitating F-coil. A LaB$_6$ cathode injects electrons from the edge of the confinement volume. The electrons follow the poloidal field lines of the coil and drift toroidally.  The beam path is visualized using dilute helium gas ($P \sim 10^{-3}$~mbar). Reflections of the red laser displacement sensors are also visible. The viewing angle is restricted by the tube of the DN40CF flange.}
    \label{fig:visualization}
\end{figure}

APEX-LD employs a lanthanum hexaboride (LaB$_6$) thermionic electron emitter mounted to a linear manipulator in the midplane of the levitation chamber (see Fig.~\ref{fig:levsys}). The crystal emission plane is oriented to emit electrons vertically, i.e., parallel to the outer magnetic field lines of the F-coil.\cite{deller_diocotron_2024} By injecting helium gas into the UHV chamber, the continuous electron beam can be imaged, a photograph of which is presented in Fig.~\ref{fig:visualization}. Visualization is typically performed with an emitter heating current of $1.2$~A, an acceleration potential of $-100$~V, and a helium pressure in the range of $10^{-5}$ to $10^{-3}$~mbar.

The $100$-eV electrons excite and ionize the helium atoms.  The resultant radiative decay produces a faint glow that illuminates the electron beam path. In addition to the gyromotion around the magnetic field (not discernible in Fig.~\ref{fig:visualization}), the electrons stream poloidally along the field lines and through the center of the F-coil, and they drift toroidally in the counter-clockwise direction (viewed from above) due to field curvature and grad-B drifts.\cite{northrop_stability_1960} The gyromotion occurs on the GHz scale, the poloidal motion on the MHz scale, and the toroidal motion on the kHz scale. The helical path of the poloidal and toroidal motion is clearly visible in Fig.~\ref{fig:visualization}.  The magnetic focusing of the beam as it enters the high-field region in the center of the F-coil is also visible. After $\sim 5$ poloidal transits, collisions and shear in the electron drift velocities causes the beam to smear into a diffuse toroidal surface. 

\subsection{\label{sec:experiments:plasma}Pure-electron plasmas}

We have performed preliminary experiments with pure-electron plasmas that verify the trapping capabilities of APEX-LD.  Under UHV conditions, the LaB$_6$ emitter on the edge of the confinement volume is switched from ground to a negative bias to produce a short pulse of electrons (filling time $100$~µs to $1$~s; injection energy $50$ to $150$~eV). The emitted electrons follow the drift orbits shown in Fig.~\ref{fig:visualization}. A small fraction diffuses radially inward and establishes a non-neutral plasma with a large space-charge potential.\cite{deller_diocotron_2024}

The dynamics of the trapped electrons are observed during the hold period, which commences after the fill is terminated by grounding the emitter bias. Plasma fluctuations are detected using capacitive wall probes installed on the equator of the confinement volume. Time-resolved Fourier spectra of the induced image currents ubiquitously exhibit a distinct mode that is consistent with a diocotron mode with a toroidal mode number $\ell = 1$.\cite{deller_diocotron_2024} The mode frequency $f_0$ is typically between $100$ and $400$~kHz, with harmonics often visible into the MHz range (these arise as a consequence of the small size of the probes relative to the plasma).\cite{singha_understanding_2025}

\begin{figure}
    \centering
    \includegraphics[width=1\linewidth]{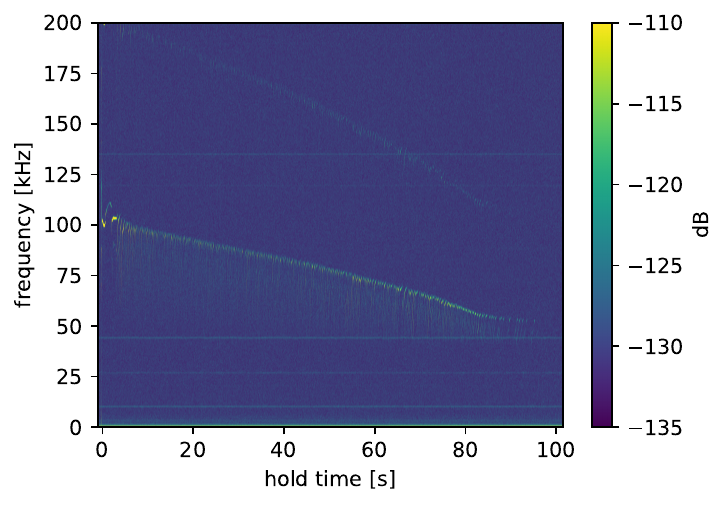}
    \caption{An example FFT spectrogram of wall-probe measurements with a trapped pure-electron plasma.  The slowly decaying toroidal mode ($f_0 \approx 100$~kHz) demonstrates long-term plasma confinement.}
    \label{fig:plasma}
\end{figure}

Low-order diocotron modes can be remarkably stable.\cite{degrassie_waves_1980} In APEX-LD, the mode commonly persists for $t_\mathrm{hold} \approx 10$~s. For certain injection parameters, however, we have observed plasma oscillations throughout much longer hold times, as demonstrated in Fig.~\ref{fig:plasma}. For this measurement, the initially stable mode quickly evolves into a series of upwards chirps. The end frequency steadily droops from roughly 100 to 50~kHz, presumably due to the loss of electrons and the corresponding decrease in space-charge potential.\cite{deller_diocotron_2024} The mode is clearly visible for $t_\mathrm{hold} \approx 90$~s, which illustrates the impressive confinement capabilities of levitating dipole traps. Even longer confinement times of pure-electron plasmas ($>300$~s) have been reported by RT-1.\cite{yoshida_magnetospheric_2010, saitoh_confinement_2010}

\section{\label{sec:discussion}Discussion and future work}

In this paper, we have presented the design and operation of APEX-LD, a new device which can repeatedly cool, inductively charge, and stably levitate a closed-loop superconducting coil under UHV conditions. Since the first charging and levitation trials in August 2023, we have performed field line visualization tests and demonstrated excellent confinement of pure-electron plasmas.\cite{deller_diocotron_2024}  The success of these measurements, coupled with the robustness of the coil and reliability of our preparation procedures, provide us with confidence that this novel device is well-suited for future experiments with pair plasmas. 

The fully soldered construction of the NI ReBCO F-coil offers unique benefits in the application of a levitated dipole. Mechanically, the solder provides remarkable robustness. We have verified this through dozens of thermal and electrical cycles, each ending with a quench, without any apparent degradation in performance. The solder provides good thermal conductivity between the HTS windings and case, which helps to minimize the preparation time. From room temperature, the system cools to $20$~K in $5.5$~h, with the F-coil trailing by $1$~h.  From its base temperature, the F-coil can be prepared for levitation in $70$~minutes (i.e, 35\% of our levitation time of $195$~minutes).  After levitation, the warming phase of preparation can be significantly shortened, allowing for a recharging time of only $45$~minutes. Accordingly, multiple cycles can be conducted per day.

The single-pancake F-coil design is a compromise necessitated by the availability of only 12-mm-wide HTS tape at the time of manufacture. Although the coil performs well (resistance at base temperature of $72$~n$\Omega$), the radial bands introduce asymmetries that we would rather avoid. In the near future, we plan to upgrade the F-coil to a double-pancake wound from a single piece of $6$-mm-wide NI ReBCO tape. In addition to enhancing the toroidal symmetry of the magnetic field, we estimate the increased inductance and reduced resistance could extend our $L/R$ decay time by an order of magnitude. Our levitation time is presently limited by exposure of the F-coil to room-temperature blackbody radiation. For this reason, an actively-cooled heat shield is under construction that will enclose over $95$\% of the surface area of the levitation region (see Fig.~\ref{fig:schematic}) and reduce the F-coil warming rate.

Confinement of a pair plasma in APEX-LD requires the injection of both positive and negative charge species; this is a challenge because the positrons arrive independently of electrons. One approach is to generate neutral, Rydberg positronium atoms, which can drift across field lines before undergoing ionization within the confinement volume.\cite{pedersen_plans_2012} However, this method must contend with compounding inefficiencies.\cite{stoneking_new_2020} Another promising injection strategy is based on drifting bunches of positrons into an established electron plasma.\cite{singer_injection_2021} Our experiments with electrons have enabled us to refine our techniques for the production and control of stable non-neutral electron plasmas in APEX-LD. Cold, dense pulses of positrons, supplied by linear traps,\cite{deller_buffer-gas_2023, singer_multi-cell_2023} will arrive from above, along the center axis (see Fig.~\ref{fig:schematic}), and will be injected using an E$\times$B drift technique.\cite{stenson_lossless_2018} This uses electrodes (incorporated into the heat shield design) to steer the charged particles from the open transport field lines into the closed confinement volume.\cite{nisl_positron_2020}

\begin{figure}
    \centering
    \includegraphics[width=1\linewidth]{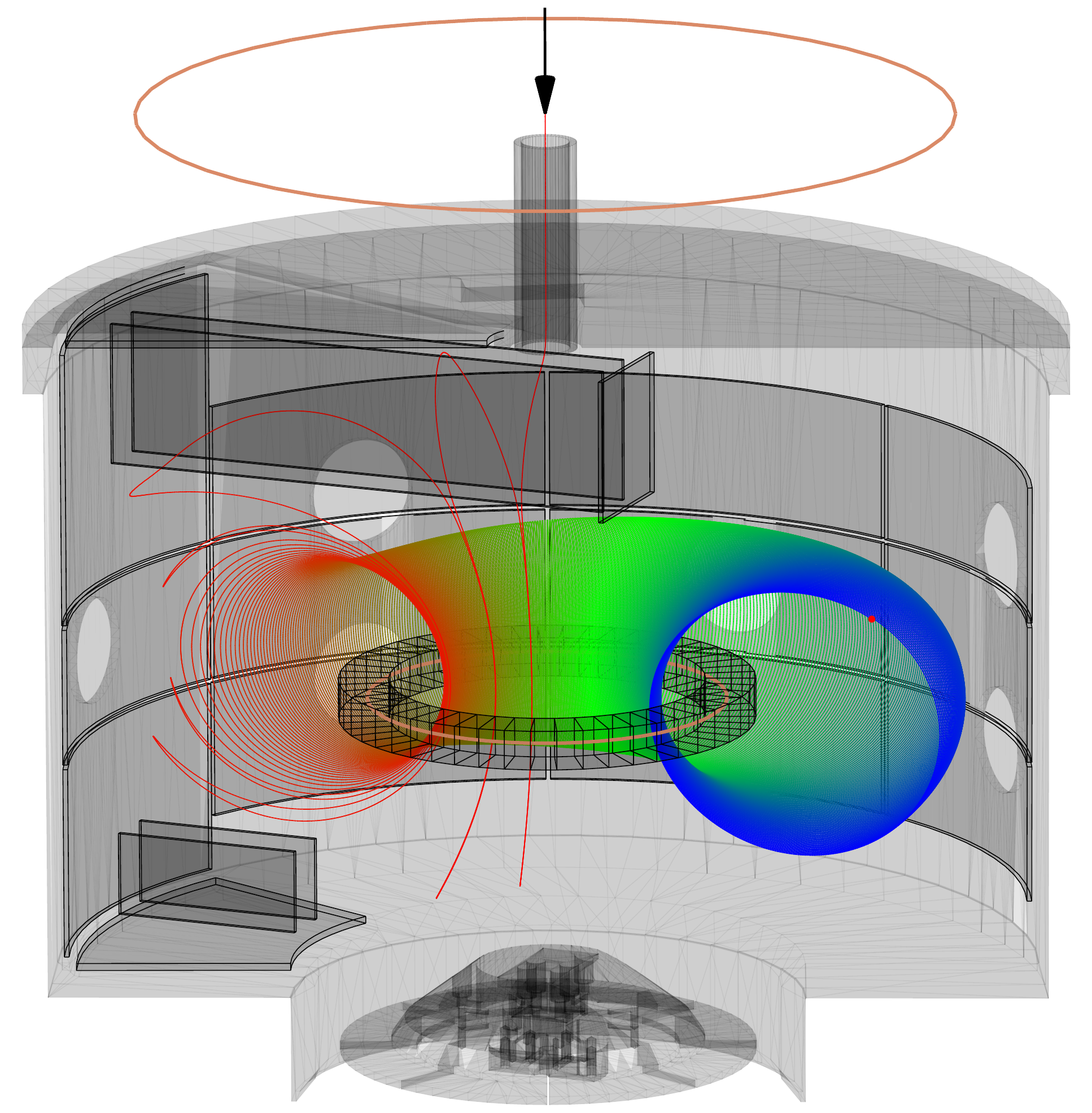}
    \caption{Simulation of injection of a positron into APEX-LD. The particle is represented by the red dot (right), and the color gradient (red-green-blue line) represents the passage of time along its trajectory. It enters from above along the cylindrical axis and passes through the E$\times$B plates, undergoing several poloidal reflections between the biased lifter/wall electrodes before drifting onto a closed field line. The injected positron drifts toroidally in the clockwise direction (viewed from above).}
    \label{fig:simulation}
\end{figure}

Figure~\ref{fig:simulation} presents a trajectory calculation of E$\times$B injection into APEX-LD. The particle enters through the top, passing first through a static, orthogonal electric field that produces an E$\times$B drift towards the closed field lines. Biased upper and outer wall electrodes prevent the particle from colliding with the thermal shield.  Simulations of positron distributions typical of linear traps suggest that high injection efficiencies are achievable.  Lossless drift injection of positron bunches into a supported dipole trap was recently demonstrated experimentally.\cite{stenson_lossless_2018, deller_injection_2024}

APEX-LD has also been designed with diagnostics for pair plasmas in mind.  The primary diagnostic for annihilation gamma-rays is an array of $48$ bismuth germanate (BGO) detectors (not shown in Fig.~\ref{fig:schematic}).  The detector arrangement will exploit the open geometry of the LD to maximize the coverage of lines of response for coincidence detection of back-to-back 511~keV gamma rays.\cite{von_der_linden_annihilation-gamma-based_2023} Additional planned diagnostics include capacitive wall probes mounted inside the heat shield, and a pulsed gas jet to instigate annihilation events.

\begin{acknowledgments}

The authors would like to thank J. Horn-Stanja, H. Saitoh, and T. Sunn Pedersen for their contributions to the project, as well as THEVA GmbH for collaboration on the coil designs. We also acknowledge the guidance of C. Hugenschmidt and R. Neu, and support from the DFG (HU 978/20-1, STE 2614/2-1), ERC (Horizon 2020, Grant 741322), NSF (PHY-2206620), DOE (DE-SC0016532), the UCSD Foundation, and the Helmholtz Association (VH-NG-1430).

\dots.
\end{acknowledgments}

\bibliographystyle{aipnum4-2}
\bibliography{lib}

\end{document}